# Auditory Brainstem Response in Infants and Children with Autism: A Meta-Analysis

Short title: Auditory Brainstem Response in Autism

Oren Miron, Andrew L. Beam, Isaac S. Kohane

Author affiliation: Department of Biomedical Informatics, Harvard Medical School, Boston, Massachusetts (OM, ALB & ISC).

Corresponding author: Oren Miron, MA, Department of Biomedical Informatics, Harvard Medical School, 10 Shattuck Street, Boston, Massachusetts 02115. Email- oren_miron@hms.harvard.edu. Phone- (617) 432-2144. Fax- (617) 432-0693.

**Acknowledgments:** We thank all the authors that created the studies that we analyzed in our meta-analysis and the participants who made those studies possible. We would like to thank Drs. Karmel, Cohen and Gardner (Cohen et al., 2013) for sending mean latencies and gender ratios from their infant autism study. None of the authors received financial compensation for providing the information for this study. The study was presented as a poster in the Neurodevelopmental Disorders Symposium by Massachusetts General Hospital, Boston Children's Hospital and MIT (Boston, MA; November 2nd 2016). All other authors report no biomedical financial interests or potential conflicts of interest. All authors have seen and approved this manuscript. This is the pre-peer reviewed version that was accepted for publication in Autism Research on October 5th 2017.

**Lay Summary:** Our analysis of previous studies showed that infants and children with autism have a slower brain response to sound, while adults have a faster brain response to sound. This suggests that slower brain response in infants may predict autism risk. Brain response to sound is routinely tested on newborns to screen hearing impairment, which has created large data to afford replication of these results.

Infants with autism were recently found to have prolonged Auditory Brainstem Response (ABR); however, at older ages, findings are contradictory. We compared ABR differences between participants with autism and controls with respect to age using a meta-analysis. Data sources included MEDLINE, EMBASE, Web of Science, Google Scholar, HOLLIS and ScienceDirect from their inception to June 2016. The 25 studies that were included had a total of 1349 participants (727 participants with autism and 622 controls) and an age range of 0-40 years. Prolongation of wave V in autism had a significant negative correlation with age ($R^2$=0.23; P=.01). The 22 studies below age 18 years showed a significantly prolonged wave V in autism (Standard Mean Difference=0.6 [95% CI, 0.5 to 0.8]; P<.001). The 3 studies above 18 years of age showed a significantly shorter wave V in autism (SMD=-0.6 [95% CI, -1.0 to -0.2]; P=.004). Prolonged ABR was consistent in infants and children with autism, suggesting it can serve as an autism biomarker at infancy. As the ABR is routinely used to screen infants for hearing impairment, the opportunity for replication studies is extensive.

The Auditory Brainstem Response (ABR) is an auditory evoked potential that is recorded through electrodes on the scalp. The evoked potential is recorded as a waveform that is characterized by 5 waves, with the first wave (wave I) originating at the auditory nerve and the fifth wave (wave V) originating at the upper brainstem (Starr, 1976). Recent publications show that wave V latency is prolonged in infants who were later diagnosed with autism (Cohen et al., 2013; Miron et al., 2015), a neurodevelopmental disorder that impairs social communication (2013). At older ages, some studies found prolonged wave V latency in autism (Azouz et al., 2014; Dabbous, 2012; Fujikawa-Brooks et al., 2010; Gillberg et al., 1983; Kwon et al., 2007; Magliaro et al., 2010; Ornitz et al., 1980; Rosenblum et al., 1980; Rosenhall et al., 2003; Roth et al., 2012; Russo et al., 2009; Sersen et al., 1990; Skoff et al., 1980; Sohmer & Student, 1978; Student & Sohmer, 1978; Tanguay et al., 1982; Tas et al., 2007; Tharpe et al., 2006; Ververi et al., 2015; Wong & Wong, 1991), while others found shorter wave V latency (Courchesne et al. 1985; Grillon et al. 1989; Rumsey et al. 1984).

Prolonged wave V in infants with autism may relate to brain overgrowth in infants with autism (Courchesne, et al., 2011; Courchesne et al., 2003; Redcay & Courchesne, 2005), since head circumference correlates with ABR latency (Mitchell et al., 1989). Brain overgrowth in infants with autism often progressed to undergrowth by adulthood (Courchesne et al., 2011; Redcay & Courchesne, 2005). This suggests that the finding of prolonged ABR in infants with autism may progress to shorter ABR in adulthood. Another finding that may relate to wave V prolongation in autism is the impaired myelination in autism (Wolff et al., 2012), which may also relate to prolonged auditory cortical responses in autism (Roberts et al., 2013).

Aside from wave V, other ABR waves also show prolongation in autism (Rosenhall et al., 2003). This meta-analysis' focus on Wave V is due to the relative consistency with which wave V appears in infant hearing screening (Driscoll & McPherson, 2010), thereby creating extensive opportunity for replication by others. If results from this meta-analysis are born out by future studies, then wave V screening as part of standard ABR of infant hearing screening could be used for earlier detection of autism. As the median age for autism diagnosis is 4.1 years (Centers for Disease Control, 2014), earlier diagnosis could lead to earlier treatment, which may result in improved outcomes, even with existing treatment modalities (Dawson et al., 2010; Silverstein & Radesky, 2016; Siu et al., 2016).

## METHODS AND MATERIALS

**Search**

We performed a literature search of MEDLINE, EMBASE, Web of Science, Google Scholar, HOLLIS and ScienceDirect from their inception to June 3$^{rd}$ 2016 (Supplementary Figure-1). The search yielded 50 studies on ABR in autism compared to controls that were published between 1975 (Ornitz & Walter, 1975) and 2015 (Miron et al., 2015).

**Study selection**

Studies were excluded if they (a) did not specify wave V latency or age values (b) were not written in English, or (c) used a sample comprised entirely of cases with a specific genetic abnormality (Figure 1 & Supplementary Table-1).

After exclusion, 25 studies remained (50% inclusion proportion), with 1349 participants (727 participants with autism and 622 controls), and an age range of 0-40 years.

**Figure 1: Inclusion of studies in meta-analysis**

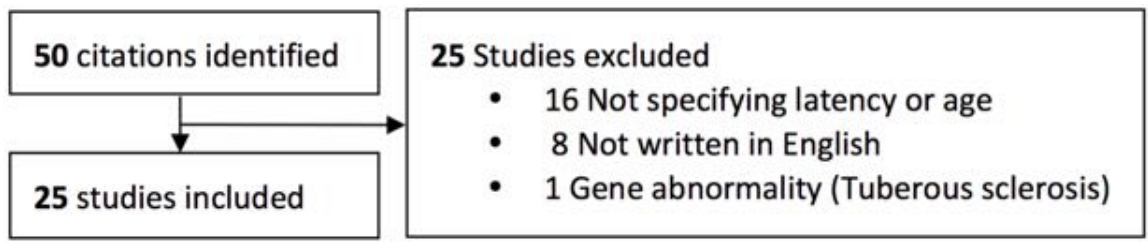

**Data Extraction and Synthesis:**

Values for the analysis were extracted from the included articles and from data sent by authors (Supplementary Table-2 and Supplementary Figure-2, 3 & 4). Data extraction and analysis was performed three times to ensure accuracy. Autism prolongation percentage was calculated as mean wave V latency in the autism group divided by mean wave V latency in the control group.

**Statistical analysis**

The analysis was performed with R statistical software (Version 3.3.1), with the packages "ggplot2" and "metafor". The effect of confounding factors on autism prolongation in wave V was examined using linear regression weighted by sample size. There was no significant correlation with male ratio, stimulus intensity, stimulus rate, sweeps amount, and publication year (Table 1).

**Table 1. Characteristics of the 25 included studies**

| Study (1st author) | Age (year) | Sample (N) | Male (%) | Intensity (dB nHL) | Rate (clicks/sec) | Sweeps (trials) | Publication (year) |
|---|---|---|---|---|---|---|---|
| Cohen† | 0.0 | 70 | 59 | 80 | 12.9 | 3,072 | 2013 |
| Miron† | 0.1 | 60 | 80 | 85 | 39.1 | 2,000 | 2015 |
| Dabbous | 2.4 | 50 | 66 | 90 | 27.5 | N/A | 2012 |
| Roth | 2.6 | 52 | 83 | 85 | 39.1 | 2,000 | 2012 |
| Kwon | 3.3 | 121 | 82 | 90 | 13 | N/A | 2007 |
| Wong | 3.4 | 129 | 72 | 80 | 10 | 2,048 | 1991 |
| Tas | 3.9 | 42 | 75 | 80 | 16 | 2,000 | 2007 |
| Ververi | 4.0 | 86 | 100 | 70 | N/A | 2,048 | 2015 |
| Azouz | 5.0 | 45 | 76 | N/A | N/A | N/A | 2014 |
| Tanguay | 5.5 | 28 | 86 | 72 | 20 | 1,500 | 1982 |
| Tharpe | 5.7 | 36 | 90 | 80 | 21.1 | 2,000 | 2006 |
| Ornitz | 6.0 | 15 | N/A | 68 | 10 | 1,024 | 1980 |
| Sohmer‡ | 7.5 | 31 | N/A | 75 | 15 | 1,024 | 1978 |
| Student‡ | 8.0 | 24 | N/A | 75 | 10 | N/A | 1978 |
| Rosenhall (female) | 9.6 | 54 | 0 | 80 | 22.5 | 1,012 | 2003 |
| Russo | 9.8 | 39 | 73 | N/A | 13 | 3,000 | 2009 |
| Rosenblum | 9.8 | 12 | 67 | 60 | 10 | 1,000 | 1980 |
| Skoff | 10.3 | 36 | 53 | 60 | 10 | 2,000 | 1980 |
| Rosenhall (male) | 10.3 | 106 | 100 | 80 | 22.5 | 1,012 | 2003 |
| Fujikawa-Brooks | 10.8 | 40 | 88 | 75 | 19 | 1,024 | 2010 |
| Gillberg | 11.3 | 55 | 65 | 80 | N/A | 2,048 | 1983 |
| Sersen | 11.5 | 83 | 100 | 50 | 10.3 | 1,500 | 1990 |
| Magliaro | 12.1 | 41 | 65 | 80 | 19 | 2,000 | 2010 |
| Rumsey | 19.4 | 50 | 92 | 80 | 11 | 2,048 | 1984 |
| Courchesne§ | 19.6 | 28 | 86 | 70 | 37 | 2,750 | 1985 |
| Grillon§ | 21.7 | 16 | 100 | 70 | 7 | 2,000 | 1989 |
| **Autism prolongation effect** | $R^2=0.006$; NS | | $R^2=0.0008$; NS | $R^2=0.1$; NS | $R^2=0.003$; NS | $R^2=0.01$; NS | |

†-Both studies used preterm infants in autism and control groups. ‡-Autism participants overlap between both studies. §-Autism participants overlap between studies. N/A-Not Available. NS- Non significant weighted linear regression (p>0.05); dB nHL- Decibels above Normal Hearing Level.,

Age correlation with autism prolongation of wave V was determined using linear regression weighted by sample size. A similar analysis was performed for waves III and I (Supplementary Figure-5) and for Inter-Peak Latencies (IPL) I-V, I-III and III-V (Supplementary Figure-6). Comparison of Standard Mean Difference (SMD) of wave V in autism vs. controls made use of random effect model with the DerSimonian and Laird method (DerSimonian & Laird, 1986). Analysis was done separately in studies with mean age below 18 years and mean age above 18 years.

## RESULTS

### Wave V autism prolongation

We first measured the effect of age on autism prolongation of wave V using a weighted linear regression model. Age had a significant negative effect on prolongation of wave V in autism ($R^2$=0.227, $F_{(1,24)}$=7.04, $P$=0.013, B=-0.23; Figure 2).

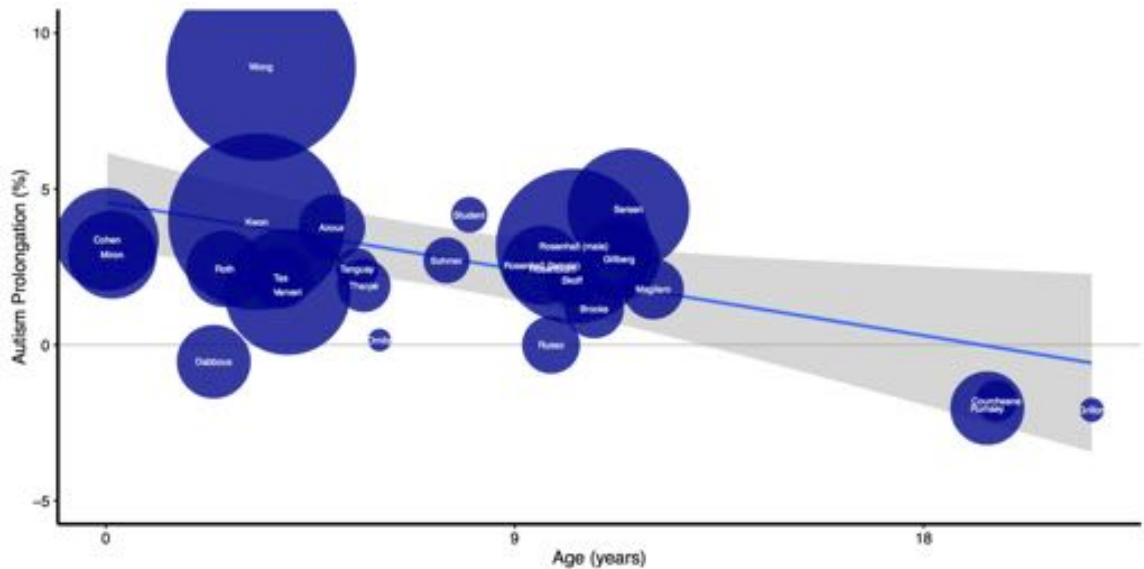

**Figure 2: Correlation of autism prolongation and age**
Legend: Y=Wave V autism prolongation in percentage. X=mean age at time of ABR in years. Blue line=Linear regression. Grey area=Linear regression confidence interval of 95%. White names indicate first author name and circle size corresponds to sample size. For example, Circle "Cohen" represents Cohen et al 2013 and the size corresponds to a sample size of 70 participants.

### Standardized Mean Difference

In order to measure the SMD before and after adulthood, a separate analysis was performed for studies with mean age below 18 years of age and above 18 years of age.

The 22 studies below 18 years of age showed significant autism prolongation of wave V (SMD=0.632, CI 0.476 to 0.788, *P*<.001). The three studies above 18 years of age showed significant autism shortening of wave V (SMD=-0.607, CI -1.021 to -0.194, *P*=.004). When examining all 25 studies together, the autism prolongation of wave V was significant (SMD=0.505, CI 0.309 to 0.700, *P*<.001; Figure 3).

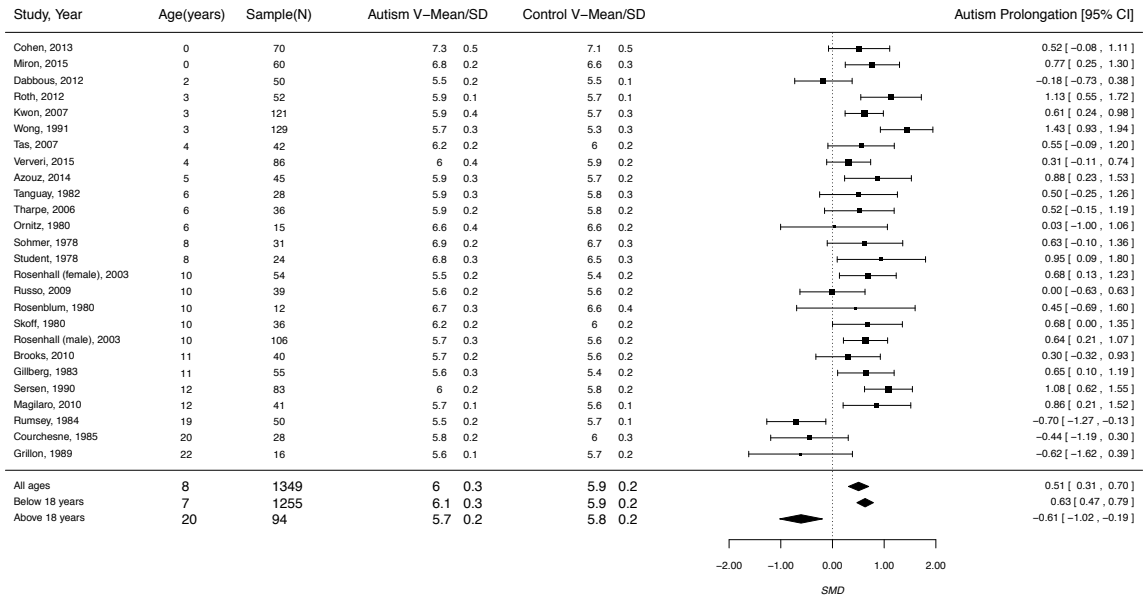

**Figure 3: Forest plot of autism prolongation**

Legend: Study (first author), year (publication); Age (mean in years); Sample (combined for autism and controls); V- Wave V latency; SD- Standard Deviation; Black square indicates SMD and error bars indicate confidence interval of 95%, which are indicated numerically under Autism Prolongation [95% CI]. Square size indicates the proportional weight of the study on the combined SMD.

## DISCUSSION

This meta-analysis found that autism prolongation has a significant negative correlation with age. Further, ABR studies below 18 years of age found wave V prolongation in autism (Cohen et al., 2013; Dabbous, 2012; Kwon et al., 2007; Miron et al., 2015; Roth et al., 2012; Wong & Wong, 1991), while studies above 18 years of age found wave V shortening (Courchesne et al., 1985; Grillon et al., 1989; Rumsey et al., 1984). Autism prolongation did not correlate significantly with study settings such as gender and stimulus intensity. The autism prolongation of infants and children was significant in wave V, which originates from the highest area of the brainstem. In this regard, it is worth noting four studies that were excluded from the meta-analysis for not specifying wave V but did specify IPL I-V (a higher part of wave V), where they found autism prolongation in a combined sample of 755 children (Maziade et al., 2000; McClelland et al. 1992; Taylor et al. 1982; Thivierge et al. 1990).

Results demonstrated a negative effect of age on autism prolongation at wave V, with prolongation in infancy and shortening at adulthood. This is similar to the negative effect of age on brain overgrowth in autism, with overgrowth in infancy and undergrowth in adulthood (Courchesne, et al., 2011; Redcay & Courchesne, 2005). These studies suggest that in the first days of life, before the increased brain growth period, there may be a smaller brain size in autism. Examining if this abnormality affects ABR requires that the post-natal analysis for the first days of life should be done separately for subsequent pediatric development. Follow-up studies to determine if the brain overgrowth and the V wave signal have any association also include: measuring the correlation between the head circumference (absolute size and growth rate) and V wave timing in both children with autism and matched controls. Furthermore, given the impaired myelination observed in autism (Wolff et al., 2012), the delayed V wave finding could be correlated with white fiber connectivity (i.e. with DTI MRI imaging). Finally, the auditory sensitivity that burdens many individuals with autism (Rosenhall et al., 1999) could also be correlated with the V wave delay.

Autism is a heterogeneous disorder with several subgroups (Doshi-Velez, et al., 2014; Karmel et al., 2010), which suggests that autism prolongation may occur in a subgroup of autism cases. This notion was supported by several studies in the meta-analysis (Miron et al., 2015; Rosenhall et al., 2003; Roth et al., 2012; Ververi et al., 2015). For example, an infant study found wave V prolongation in 70% of autism and 20% of controls (Miron et al., 2015). The studies that were analyzed used different prolongation thresholds, which prevented a subgroup comparison in the meta-analysis. However, this heterogeneity does suggest the opportunity in studying the genetics of the individuals identified with prolonged V wave. For example, assessing the genes involved in myelination and with variants previously associated with autism. These include X-linked dystrophin-related protein 2 gene (DRP2) which regulates Schwann cell myelination and in which loss-of-function mutations have been found in patients with autism (Toma et al., 2014), large non-coding RNA mutations associated with delayed myelination and autism (Talkowski et al., 2012), and mutations in ERBB4 associated with autism and a tyrosine receptor kinase regulation of myelination (Gai et al., 2012). However, rather than focusing on those specific genes, a genome-wide approach is now feasible and affordable but it would require the reconsent and recontact of those families in which autism and prolonged ABR V waves were found.

However, even without determining the specific mechanisms by which the prolonged V wave is associated with autism, if it is found on replication studies to have sufficient specificity it could be used as a very low cost biomarker that fits into current pediatric practice. Therefore the only additional requirement would be the application of signal processing algorithms well within the capabilities of commodity consumer computing.

**References**

A P A. (2013). Diagnostic and statistical manual of mental disorders (DSM-5®).

Azouz, H. G., Kozou, H., Khalil, M., Abdou, R. M., & Sakr, M. (2014). The correlation between central auditory processing in autistic children and their language processing abilities. *International Journal of Pediatric Otorhinolaryngology*, *78*(12), 2297–2300. doi:10.1016/j.ijporl.2014.10.039

Cohen, I. L., Gardner, J. M., Karmel, B. Z., Phan, H. T., Kittler, P., Gomez, T. R., … Barone, A. (2013). Neonatal brainstem function and 4-month arousal-modulated attention are jointly associated with autism. *Autism Research : Official Journal of the International Society for Autism Research*, *6*(1), 11–22. doi:10.1002/aur.1259

Courchesne, E., Campbell, K., & Solso, S. (2011). Brain growth across the life span in autism: age-specific changes in anatomical pathology. *Brain Research*, *1380*, 138–145. doi:10.1016/j.brainres.2010.09.101

Courchesne, E., Carper, R., & Akshoomoff, N. (2003). Evidence of brain overgrowth in the first year of life in autism. *The Journal of the American Medical Association*, *290*(3), 337–344. doi:10.1001/jama.290.3.337

Courchesne, E., Courchesne, R. Y., Hicks, G., & Lincoln, A. J. (1985). Functioning of the brain-stem auditory pathway in non-retarded autistic individuals. *Electroencephalography and Clinical Neurophysiology*, *61*(6), 491–501.

Courchesne, E., Mouton, P. R., Calhoun, M. E., Semendeferi, K., Ahrens-Barbeau, C., Hallet, M. J., … Pierce, K. (2011). Neuron number and size in prefrontal cortex of children with autism. *The Journal of the American Medical Association*, *306*(18), 2001–2010. doi:10.1001/jama.2011.1638

Dabbous, A. O. (2012). Characteristics of auditory brainstem response latencies in children with autism spectrum disorders. *Audiological Medicine*, *10*(3), 122–131. doi:10.3109/1651386X.2012.708986

Dawson, G., Rogers, S., Munson, J., Smith, M., Winter, J., & et al. (2010). Randomized, controlled trial of an intervention for toddlers with autism: the Early Start Denver Model. *Pediatrics*, *125*(1), 17–23.

DerSimonian, R., & Laird, N. (1986). Meta-analysis in clinical trials. *Controlled Clinical Trials*, *7*(3), 177–188. doi:10.1016/0197-2456(86)90046-2

Developmental Disabilities Monitoring Network Surveillance Year 2010 Principal Investigators, & Centers for Disease Control and Prevention (CDC). (2014). Prevalence of autism spectrum disorder among children aged 8 years - autism and developmental disabilities monitoring network, 11 sites, United States, 2010. *MMWR. Surveillance Summaries : Morbidity and Mortality Weekly Report. Surveillance Summaries / CDC*, *63*(2), 1–21.

Doshi-Velez, F., Ge, Y., & Kohane, I. (2014). Comorbidity clusters in autism spectrum disorders: an electronic health record time-series analysis. *Pediatrics*, *133*(1), e54–63. doi:10.1542/peds.2013-0819

Driscoll, C. J., & McPherson, B. (2010). *Newborn Screening Systems: The Complete Perspective* (reprint.). Plural Publishing.

Fujikawa-Brooks, S., Isenberg, A. L., Osann, K., Spence, M. A., & Gage, N. M. (2010).

The effect of rate stress on the auditory brainstem response in autism: a preliminary report. *International Journal of Audiology*, *49*(2), 129–140. doi:10.3109/14992020903289790

Gai, X., Xie, H. M., Perin, J. C., Takahashi, N., Murphy, K., Wenocur, A. S., … White, P. S. (2012). Rare structural variation of synapse and neurotransmission genes in autism. *Molecular Psychiatry*, *17*(4), 402–411. doi:10.1038/mp.2011.10

Gillberg, C., Rosenhall, U., & Johansson, E. (1983). Auditory brainstem responses in childhood psychosis. *Journal of Autism and Developmental Disorders*, *13*(2), 181–195.

Grillon, C., Courchesne, E., & Akshoomoff, N. (1989). Brainstem and middle latency auditory evoked potentials in autism and developmental language disorder. *Journal of Autism and Developmental Disorders*, *19*(2), 255–269. doi:10.1007/BF02211845

Karmel, B. Z., Gardner, J. M., Meade, L. S., Cohen, I. L., London, E., Flory, M. J., … Harin, A. (2010). Early medical and behavioral characteristics of NICU infants later classified with ASD. *Pediatrics*, *126*(3), 457–467. doi:10.1542/peds.2009-2680

Kwon, S., Kim, J., Choe, B. H., Ko, C., & Park, S. (2007). Electrophysiologic assessment of central auditory processing by auditory brainstem responses in children with autism spectrum disorders. *Journal of Korean Medical Science*, *22*(4), 656–659. doi:10.3346/jkms.2007.22.4.656

Magliaro, F. C., Scheuer, C. I., Assumpção Júnior, F. B., & Matas, C. G. (2010). Study of auditory evoked potentials in autism. *Pro-Fono : Revista de Atualizacao Cientifica*, *22*(1), 31–36.

Maziade, M., Mérette, C., Cayer, M., Roy, M. A., Szatmari, P., Côté, R., & Thivierge, J. (2000). Prolongation of brainstem auditory-evoked responses in autistic probands and their unaffected relatives. *Archives of General Psychiatry*, *57*(11), 1077–1083.

McClelland, R. J., Eyre, D. G., Watson, D., Calvert, G. J., & Sherrard, E. (1992). Central conduction time in childhood autism. *The British Journal of Psychiatry*, *160*, 659–663.

Miron, O., Ari-Even Roth, D., Gabis, L. V., Henkin, Y., Shefer, S., Dinstein, I., & Geva, R. (2015). Prolonged auditory brainstem responses in infants with autism. *Autism Research : Official Journal of the International Society for Autism Research*. doi:10.1002/aur.1561

Mitchell, C., Phillips, D. S., & Trune, D. R. (1989). Variables affecting the auditory brainstem response: audiogram, age, gender and head size. *Hearing Research*, *40*(1-2), 75–85.

Ornitz, E. M., Mo, A., Olson, S. T., & Walter, D. O. (1980). Influence of click sound pressure direction on brainstem responses in children. *Audiology*, *19*(3), 245–254.

Ornitz, E. M., & Walter, D. O. (1975). The effect of sound pressure waveform on human brain stem auditory evoked responses. *Brain Research*, *92*(3), 490–498.

Redcay, E., & Courchesne, E. (2005). When is the brain enlarged in autism? A meta-analysis of all brain size reports. *Biological Psychiatry*, *58*(1), 1–9.

doi:10.1016/j.biopsych.2005.03.026

Roberts, T. P., Lanza, M. R., Dell, J., Qasmieh, S., Hines, K., Blaskey, L., … Berman, J. I. (2013). Maturational differences in thalamocortical white matter microstructure and auditory evoked response latencies in autism spectrum disorders. *Brain Research*, *1537*, 79–85. doi:10.1016/j.brainres.2013.09.011

Rosenblum, S. M., Arick, J. R., Krug, D. A., Stubbs, E. G., Young, N. B., & Pelson, R. O. (1980). Auditory brainstem evoked responses in autistic children. *Journal of Autism and Developmental Disorders*, *10*(2), 215–225.

Rosenhall, U., Nordin, V., Brantberg, K., & Gillberg, C. (2003). Autism and auditory brain stem responses. *Ear and Hearing*, *24*(3), 206–214. doi:10.1097/01.AUD.0000069326.11466.7E

Rosenhall, U., Nordin, V., Sandström, M., Ahlsen, G., & et al. (1999). Autism and hearing loss. *Journal of Autism and Developmental Disorders*, *29*(5), 349–357.

Roth, D. A., Muchnik, C., Shabtai, E., Hildesheimer, M., & Henkin, Y. (2012). Evidence for atypical auditory brainstem responses in young children with suspected autism spectrum disorders. *Developmental Medicine and Child Neurology*, *54*(1), 23–29. doi:10.1111/j.1469-8749.2011.04149.x

Rumsey, J. M., Grimes, A. M., Pikus, A. M., Duara, R., & Ismond, D. R. (1984). Auditory brainstem responses in pervasive developmental disorders. *Biological Psychiatry*, *19*(10), 1403–1418.

Russo, N., Nicol, T., Trommer, B., Zecker, S., & Kraus, N. (2009). Brainstem transcription of speech is disrupted in children with autism spectrum disorders. *Developmental Science*, *12*(4), 557–567. doi:10.1111/j.1467-7687.2008.00790.x

Sersen, E. A., Heaney, G., Clausen, J., Belser, R., & Rainbow, S. (1990). Brainstem auditory-evoked responses with and without sedation in autism and Down's syndrome. *Biological Psychiatry*, *27*(8), 834–840.

Silverstein, M., & Radesky, J. (2016). Embrace the Complexity. *The Journal of the American Medical Association*, *315*(7), 661. doi:10.1001/jama.2016.0051

Siu, A. L., US Preventive Services Task Force (USPSTF), Bibbins-Domingo, K., Grossman, D. C., Baumann, L. C., Davidson, K. W., … Pignone, M. P. (2016). Screening for Autism Spectrum Disorder in Young Children: US Preventive Services Task Force Recommendation Statement. *The Journal of the American Medical Association*, *315*(7), 691–696. doi:10.1001/jama.2016.0018

Skoff, B. F., Mirsky, A. F., & Turner, D. (1980). Prolonged brainstem transmission time in autism. *Psychiatry Research*, *2*(2), 157–166.

Sohmer, H., & Student, M. (1978). Auditory nerve and brain-stem evoked responses in normal, autistic, minimal brain dysfunction and psychomotor retarded children. *Electroencephalography and Clinical Neurophysiology*, *44*(3), 380–388.

Starr, A. (1976). Correlation between confirmed sites of neurological lesions and abnormalities of far-field auditory brainstem responses. *Electroencephalography and Clinical Neurophysiology*, *41*(6), 595–608. doi:10.1016/0013-4694(76)90005-5

Student, M., & Sohmer, H. (1978). Evidence from auditory nerve and brainstem evoked responses for an organic brain lesion in children with autistic traits.

*Journal of Autism and Childhood Schizophrenia*, *8*(1), 13–20.
Talkowski, M. E., Maussion, G., Crapper, L., Rosenfeld, J. A., Blumenthal, I., Hanscom, C., ... Ernst, C. (2012). Disruption of a large intergenic noncoding RNA in subjects with neurodevelopmental disabilities. *American Journal of Human Genetics*, *91*(6), 1128–1134. doi:10.1016/j.ajhg.2012.10.016
Tanguay, P. E., Edwards, R. M., Buchwald, J., Schwafel, J., & Allen, V. (1982). Auditory brainstem evoked responses in autistic children. *Archives of General Psychiatry*, *39*(2), 174–180.
Tas, A., Yagiz, R., Tas, M., Esme, M., Uzun, C., & Karasalihoglu, A. R. (2007). Evaluation of hearing in children with autism by using TEOAE and ABR. *Autism: The International Journal of Research and Practice*, *11*(1), 73–79. doi:10.1177/1362361307070908
Taylor, M. J., Rosenblatt, B., & Linschoten, L. (1982). Auditory brainstem response abnormalities in autistic children. *The Canadian Journal of Neurological Sciences. Le Journal Canadien Des Sciences Neurologiques*, *9*(4), 429–433.
Tharpe, A. M., Bess, F. H., Sladen, D. P., Schissel, H., Couch, S., & Schery, T. (2006). Auditory characteristics of children with autism. *Ear and Hearing*, *27*(4), 430–441. doi:10.1097/01.aud.0000224981.60575.d8
Thivierge, J., Bédard, C., Côté, R., & Maziade, M. (1990). Brainstem auditory evoked response and subcortical abnormalities in autism. *The American Journal of Psychiatry*, *147*(12), 1609–1613. doi:10.1176/ajp.147.12.1609
Toma, C., Torrico, B., Hervás, A., Valdés-Mas, R., Tristán-Noguero, A., Padillo, V., ... Cormand, B. (2014). Exome sequencing in multiplex autism families suggests a major role for heterozygous truncating mutations. *Molecular Psychiatry*, *19*(7), 784–790. doi:10.1038/mp.2013.106
Ververi, A., Vargiami, E., Papadopoulou, V., Tryfonas, D., & Zafeiriou, D. (2015). Brainstem Auditory Evoked Potentials in Boys with Autism: Still Searching for the Hidden Truth. *Iranian Journal of Child Neurology*, *9*(2), 21–28.
Wolff, J. J., Gu, H., Gerig, G., Elison, J. T., Styner, M., Gouttard, S., ... IBIS Network. (2012). Differences in white matter fiber tract development present from 6 to 24 months in infants with autism. *The American Journal of Psychiatry*, *169*(6), 589–600. doi:10.1176/appi.ajp.2011.11091447
Wong, V., & Wong, S. N. (1991). Brainstem auditory evoked potential study in children with autistic disorder. *Journal of Autism and Developmental Disorders*, *21*(3), 329–340.

# Online Supplementary material

## Supplementary figure 1: Search procedure

The search was performed by the first author (OM), and was assisted by a librarian at Countway Library of Medicine (Boston, United States). The search utilized the software of the respective databases, and also made use of hand searching. The search terms used were:

MEDLINE
 ("Autistic Disorder"[Mesh] OR autis*[tiab](
AnD
("Evoked Potentials, Auditory, Brain Stem"[Mesh] OR ((auditory brainstem[tiab] OR auditory brain stem[tiab] OR auditory evoked[tiab]) AND (response*[tiab] OR potenial*[tiab])))

Embase
 ('autism'/exp OR autis*:ab,ti)
AND
('evoked brain stem auditory response'/exp OR ('auditory brainstem' NEAR/1 (response OR potential)):ab,ti
OR ((brainstem OR 'brain stem') NEAR/1 'auditory evoked'):ab,ti)

Web of Science
TS="autis*"
AND
TS=(("auditory brainstem" NEAR/1 ("response" OR "potential")) OR (("brainstem" OR "brain stem") NEAR/1 "auditory evoked"))

Google Scholar, HOLLIS and ScienceDirect

"Auditory Brainstem Response Autism"

Supplementary Table 1: Exclusion chart

| Number | Study | Title | Reason for exclusion |
|---|---|---|---|
| 1 | Maziade et al 2000(1) | Prolongation of brainstem auditory-evoked responses in autistic probands and their unaffected relatives | No-V |
| 2 | Taylor et al 1982(2) | Auditory brainstem response abnormalities in autistic children | No-V |
| 3 | Mclelland et al 1992(3) | Auditory brainstem response screening for hearing loss in high risk neonates | No-V |
| 4 | Thivierge et al 1990(4) | Brain-stem auditory evoked response (BAER): Normative study in children and adults | No-V |
| 5 | Novick et al 1980(5) | An electrophysiologic indication of auditory processing defects in autism | No-V |
| 6 | Skoff et al 1986(6) | Brainstem auditory evoked potentials in autism | No-V |
| 7 | Gillberg et al 1987(7) | Neurobiological Findings In 20 Relatively Gifted Children With Kanner-Type Autism Or Asperger Syndrome | No-V |
| 8 | Romero et al 2014(8) | AUDIOLOGIC AND ELECTROPHYSIOLOGIC EVALUATION IN THE AUTISTIC SPECTRUM DISORDER | No-V |
| 9 | Ho et al 1999(9) | Pervasive Developmental Delay in Children Presenting As Possible Hearing Loss | No-V |
| 10 | Demopoulos et al 2015(10) | Audiometric Profiles in Autism Spectrum Disorders: Does Subclinical Hearing Loss Impact Communication? | No-V |
| 11 | Kallstrand et al 2010(11) | Abnormal auditory forward masking pattern in the brainstem response of individuals with Asperger syndrome | No-V |
| 12 | Matas et al 2009(12) | Audiologic and electrophysiologic evaluation in children with psychiatric disorders | No-V |
| 13 | Fein et al 1981(13) | Clinical Correlates of Brainstem Dysfunction in Autistic Children | No-V |
| 14 | Steffenburg et al 1991(14) | Neuropsychiatric assessment of children with autism: a population-based study. | No-V |
| 15 | Yan Hua et al 2012(15) | Auditory abnormalities in children with autism.(Report) | No-V |
| 16 | Al-Ayadhi(16) | Auditory brainstem evoked response in autistic children in central Saudi Arabia | No-age |
| 17 | Zhang et al 2008(17) | Observation of the Brainstem Auditory Evoked Potentials in Autism Childhood | No-English |
| 18 | Zhang et al 2010(18) | Discussion on early diagnosis of brainstem auditory evoked potentials in Childhood Autism | No-English |
| 19 | Garruae et al 1984(19) | Auditory brain-stem evoked responses in normal and autistic children | No-English |
| 20 | Germano et al 2006(20) | Neurobiology of autism: Study of a sample of autistic children | No-English |
| 21 | Wu et al 2014(21) | The clinical analysis of results of brainstem auditory evoked potentials in 27 autism children | No-English |
| 22 | Wang et al 2009(22) | Auditory brainstem response and DPOAE in autism children | No-English |
| 23 | Wang et al 2010(23) | The Hearing Status and the Functions of Efferent System in Autistic Children | No-English |
| 24 | Rongqin Wu et al 2011(24) | The clinical applications of auditory brainstem response in juvenile with schizophrenia and autistism | No-English |
| 25 | Seri et al 1991(25) | Autism in tuberous sclerosis: evoked potential evidence for a deficit in auditory sensory processing | Tuberous sclerosis |

Supplementary Table 2: Contacting authors

An attempt was made to reach authors of studies that did not specify wave V latency or age, which is needed for the comparison between the studies (authors whose article provided sufficient data were not contacted). Of the authors that were contacted, the authors of three articles that published in 2014(26), 2013(27) and 2009(28) responded with the data (see table). The authors of older publications that were contacted did not have the data or did not respond. The author of a poster presentation from 1986(6) was contacted for more information and he replied that it was not saved, so only journal articles were used in the final analysis. An attempt was made to contact authors who did not publish in English to ask for an English version but in the 3 attempts that were made, no response was reached (18, 20, 29), so no further attempts were made.

| Corresponding Author | Response |
|---|---|
| **Dr. Rania Abdou**(26) | Sent means |
| **Dr. Ira Cohen**(27) | Sent means and gender ratio |
| **Dr. Nina Kraus**(28) | Sent means |
| **Dr. Haim Sohmer**(30) | Confirmed that his article and Student et al 1978(31) probably used the same children with autism |
| **Dr. Novick**(5) | Attempting to find data |
| **Dr. Maziade**(1) | Data not saved |
| **Dr. Thivierge**(4) | Data not saved |
| **Dr. Taylor**(2) | Data not saved |
| **Dr. McClelland**(3) | Data not saved |
| **Dr. Skoff**(6) | Data not saved |
| **Dr. Matas**(12) | No reply |
| **Dr. Al-Ayadhi**(16) | No reply |
| **Dr. Zhang**(18) | No reply |
| **Dr. Gerraue**(19) | No reply |
| **Dr. Germano**(20) | No reply |
| **Dr. Frizzo**(8) | No reply |
| **Dr. Fein**(13) | No reply |

Supplementary Figure 2: Extracting data from plot images
For 3 articles that specified mean latencies only in plot images, the values were extracted from the image (Web-Plot-Digitizer http://arohatgi.info/WebPlotDigitizer/). Extracted values were later rechecked manually. Red dots indicate the extracted values of means and standard deviations.

Rosenhall et al 2003(32)

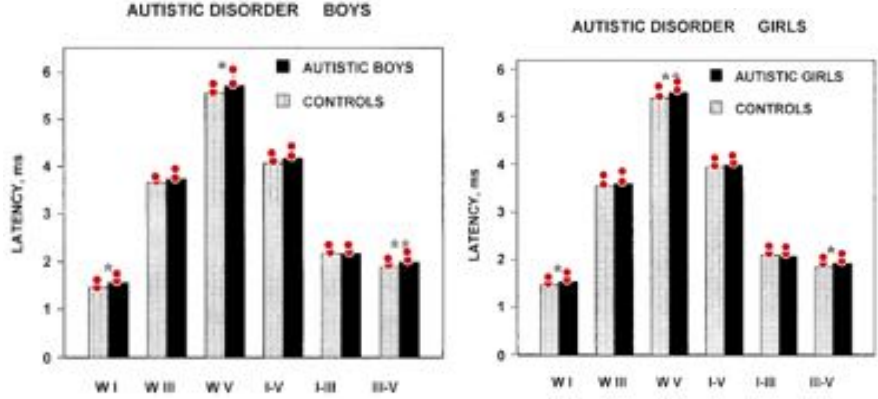

Tanguay et al 1982(33)

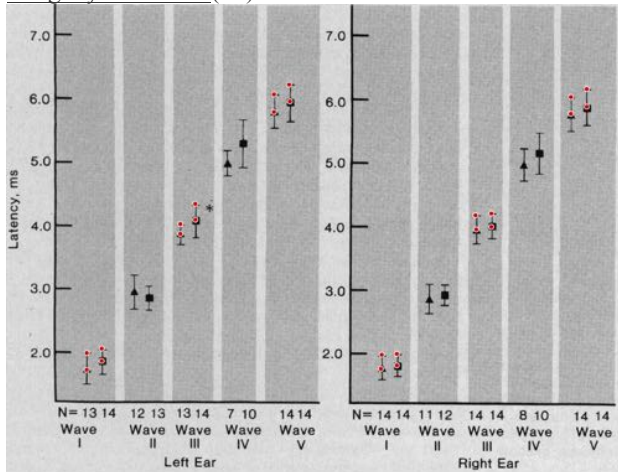

Courchesne et al 1985(34)

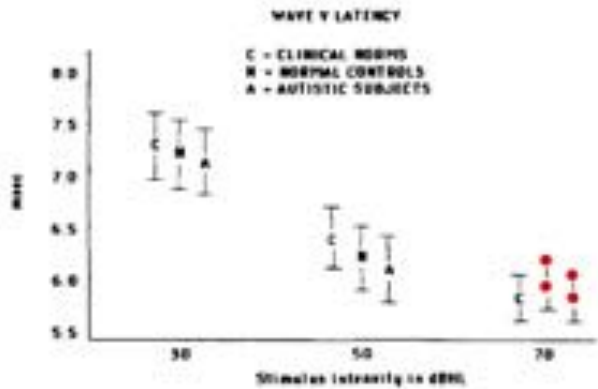

Supplementary Figure 3: Funnel plot estimation of publication bias

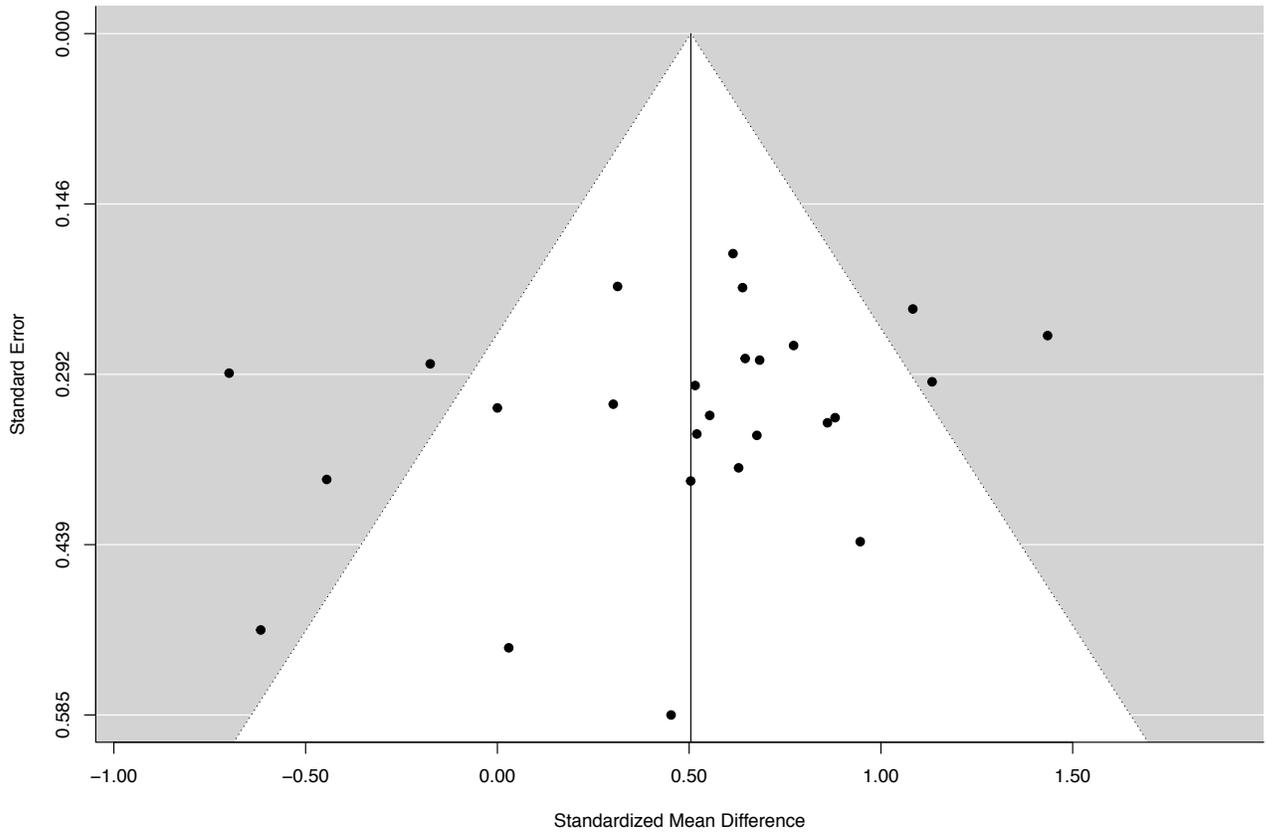

Legend: Y=Standard error of wave V autism prolongation. X=Standardized Mean Difference of wave V autism prolongation.

Supplementary Figure 4: Article comparison explanation

When articles used several sounds intensities, the analysis included the values from the intensities that were closest to the intensities of the other articles.(33, 34) One article specified all the waves of a certain intensity and only one wave for other intensities, which lead to the inclusion of the more detailed intensity.(40) Most articles specified latency for both ears, so in the few cases specified for left and/or right, the latencies were averaged. In one study a subsample of toddlers was compared to clinical norms based on young adults, so this subsample was not include in the analysis to avoid bias by the large age difference.(41) In studies using both an autism group and mild/sub-clinical autism group, the autism group was included.(42, 43) Average age was calculated by averaging the control average age and autism average age. In the few studies where average was not specified but range was, the minimum age and the maximum age were averaged. (30, 31, 44) Similarly, in the few studies where number of sweeps or rate was given as minimum and maximum values, the minimum and the maximum values were averaged.(30, 32, 34, 45) Comparison of hearing threshold exclusion was not possible due to studies using different thresholds. The majority of studies excluded abnormal hearing cases.(26, 27, 32–34, 40–43, 46–56) In cases where some latencies were not specified but could be calculated by the other latencies that were specified, such calculation was performed. For example, when IPL I-V was not specified but the absolute latencies of waves V and I were specified, the absolute latencies were used to calculate the IPL.

Supplementary Figure 5: Autism prolongation in waves III and I

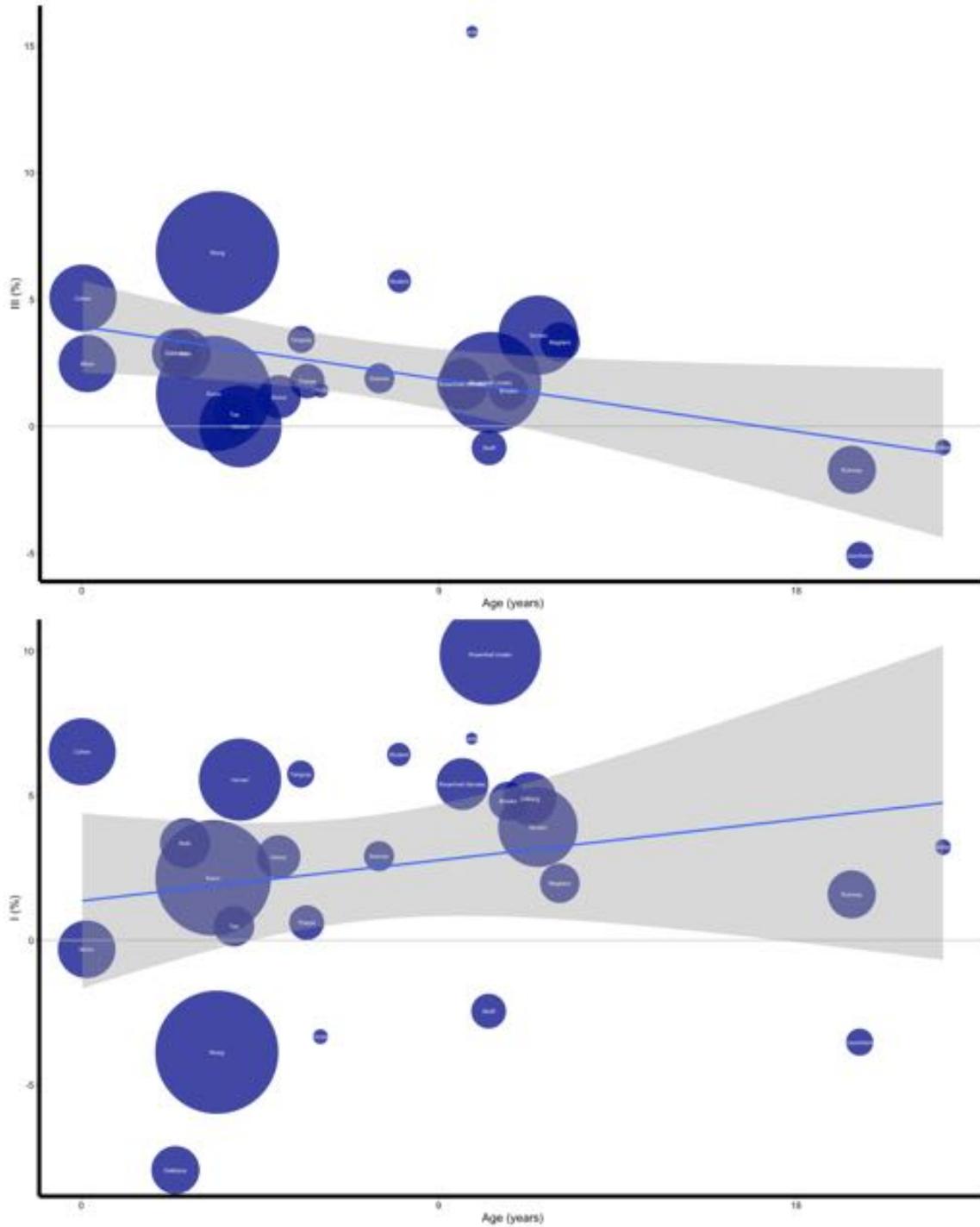

Legend: Y=Wave prolongation in percentage (III top & I bottom). X=mean age at time of ABR in years. Blue line=Linear regression. Grey area=Linear regression confidence interval of 95%. White names indicate first author name and circle size corresponds to sample size. For example, Circle "Cohen" represents Cohen et al 2013 and the size corresponds to a sample size of 70 participants. Wave III originates from a higher area compared to wave I.

Supplementary Figure 6: Autism prolongation in Inter-Peak Latencies I-V, I-III and III-V

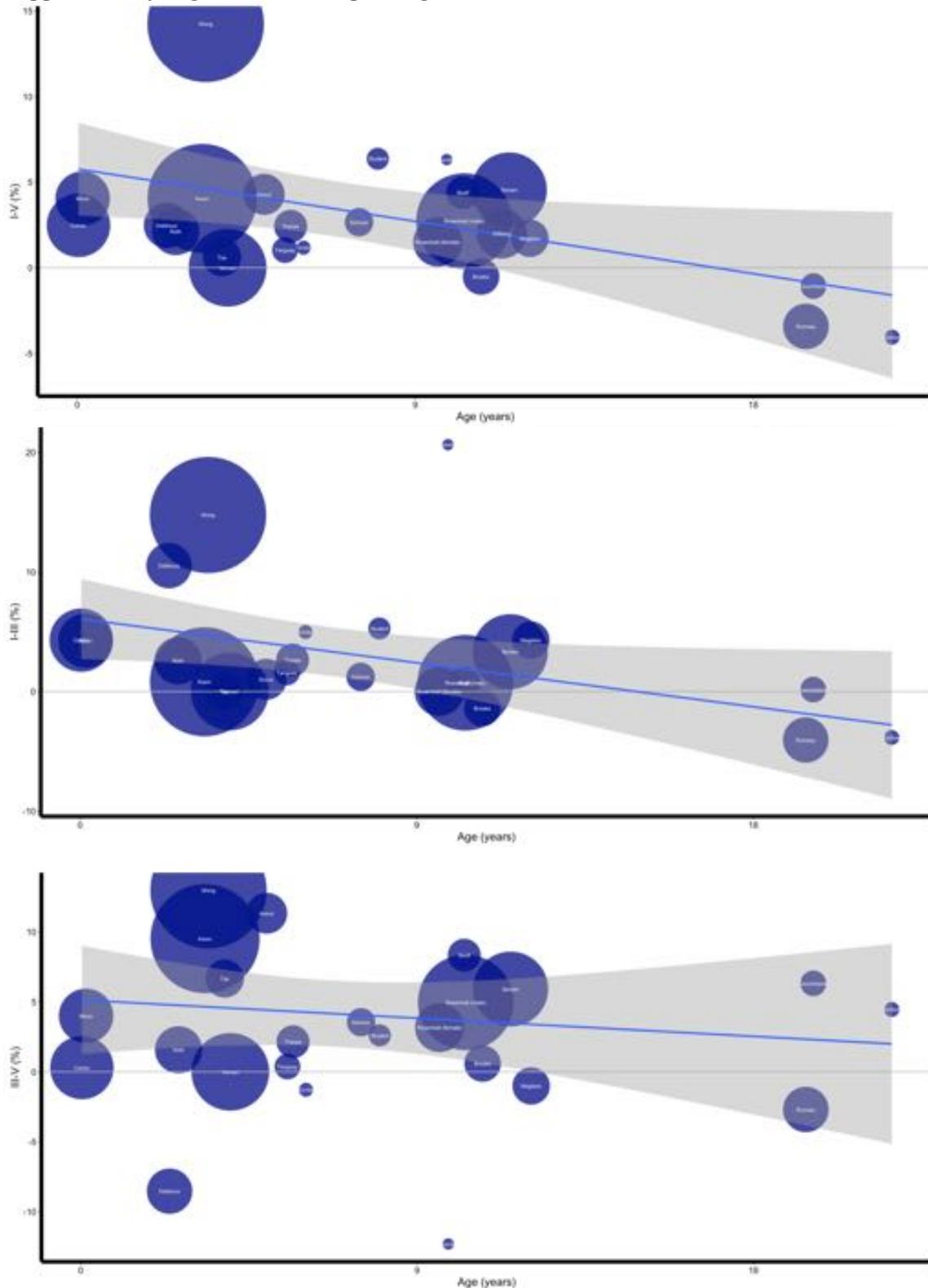

Legend: Y=Wave prolongation in percentage (I-V top, I-III middle & III-V bottom). X=mean age at time of ABR in years. Blue line=Linear regression. Grey area=Linear regression confidence interval of 95%. White names indicate first author name and circle size corresponds to sample size. For example, Circle "Cohen" represents Cohen et al 2013 and the size corresponds to a sample size of 70 participants. Inter-Peak latencies are measured as the latency difference between the absolute latency of wave I, III

**SUPPLEMENTAL REFERENCES**

1. Maziade M, Mérette C, Cayer M, Roy MA, Szatmari P, Côté R, *et al.* (2000): Prolongation of brainstem auditory-evoked responses in autistic probands and their unaffected relatives. *Arch Gen Psychiatry*. 57(11): 1077–83.
2. Taylor MJ, Rosenblatt B, Linschoten L (1982): Auditory brainstem response abnormalities in autistic children. *Can J Neurol Sci*. 9(4): 429–33.
3. McClelland RJ, Eyre DG, Watson D, Calvert GJ, Sherrard E (1992): Central conduction time in childhood autism. *Br J Psychiatry*. 160: 659–63.
4. Thivierge J, Bédard C, Côté R, Maziade M (1990): Brainstem auditory evoked response and subcortical abnormalities in autism. *Am J Psychiatry*. 147(12): 1609–13.
5. Novick B, Vaughan HG, Kurtzberg D, Simson R (1980): An electrophysiologic indication of auditory processing defects in autism. *Psychiatry Res*. 3(1): 107–14.
6. Skoff BD, Fein D, McNally B, Lucci D, Humes-Bartlo M, Waterhouse L (1986): Brainstem auditory evoked potentials in autism. *Psychophysiology*
7. Gillberg C, Steffenburg S, Jakobsson G (1987): Neurobiological findings in 20 relatively gifted children with kanner-type autism or asperger syndrome. *Dev Med Child Neurol*. 29(5): 641–49.
8. Romero A, Gução A, Delecrode C, Cardoso A, Misquiatti A, Frizzo A (2014): Audiologic and electrophysiologic evaluation in the autistic spectrum disorder. *Rev. CEFAC*. 16(3): 707–14.
9. Ho PT, Keller JL, Berg AL, Cargan AL, Haddad J (1999): Pervasive developmental delay in children presenting as possible hearing loss. *Laryngoscope*. 109(1): 129–35.
10. Demopoulos C, Lewine JD (2016): Audiometric profiles in autism spectrum disorders: does subclinical hearing loss impact communication? *Autism Res*. 9(1): 107–20.
11. Källstrand J, Olsson O, Nehlstedt SF, Sköld ML, Nielzén S (2010): Abnormal auditory forward masking pattern in the brainstem response of individuals with asperger syndrome. *Neuropsychiatr Dis Treat*. 6: 289–96.
12. Matas CG, Gonçalves IC, Magliaro FC (2009): Audiologic and electrophysiologic evaluation in children with psychiatric disorders. *Braz J Otorhinolaryngol*. 75(1): 130–38.
13. Fein D, Skoff B, Mirsky AF (1981): Clinical correlates of brainstem dysfunction in autistic children. *J Autism Dev Disord*. 11(3): 303–15.
14. Steffenburg S (1991): Neuropsychiatric assessment of children with autism: a population-based study. *Dev Med Child Neurol*. 33(6): 495–511.
15. Hua TY, Yan XC, Ping JS, Xin SB, Bo WL, Lin W (2012): Auditory abnormalities in children with autism.
16. Al-Ayadhi LY (2008): Auditory brainstem evoked response in autistic children in central saudi arabia. *Neurosciences (Riyadh)*. 13(2): 192–93.
17. Zhang Nan, Li Yuru (2008): Observation of the brainstem auditory evoked potentials in autism childhood.
18. Zhang Y, Shen H, He D (2010): Discussion on early diagnosis of brainstem

auditory evoked potentials in childhood autism. *National Medical Frontiers of China*. 5(2):
19. Garreau B, Tanguay P, Roux S, Lelord G (1984): [brain stem auditory evoked potentials in the normal and autistic child]. *Rev Electroencephalogr Neurophysiol Clin*. 14(1): 25–31.
20. Germanò E, Gagliano A, Magazù A, Calarese T, Calabrò ME, Bonsignore M, *et al. (2006): [neurobiology of autism: study of a sample of autistic children]. Minerva Pediatr*. 58(2): 109–20.
21. WU Xiao-qing, Chang He, WU Min, YAN Dong一mei (2014): The clinical analysis on results of brainstem auditory evoked potentials in 27 autism children. *Proceeding of Clinical Medicine*
22. Wang Sufang, Dong Xuelei, Wang Yongsheng (2009): Auditory brainstem response and dpoae in autism children.
23. Wang Chenrong, Hua Qingquan, Huang Zhiwu, Li Dan (2010): The hearing status and the functions of efferent system in autistic children.
24. Rongqin Wu, Shaojin Zhang, Guangqi Zhang, Chong Chen (2011): The clinical applications of auditory brainstem response in juvenile with schizophrenia and autistism.
25. Seri S, Cerquiglini A, Pisani F, Curatolo P (1999): Autism in tuberous sclerosis: evoked potential evidence for a deficit in auditory sensory processing. *Clin Neurophysiol*. 110(10): 1825–30.
26. Azouz HG, Kozou H, Khalil M, Abdou RM, Sakr M (2014): The correlation between central auditory processing in autistic children and their language processing abilities. *Int J Pediatr Otorhinolaryngol*. 78(12): 2297–2300.
27. Cohen IL, Gardner JM, Karmel BZ, Phan HT, Kittler P, Gomez TR, *et al. (2013): Neonatal brainstem function and 4-month arousal-modulated attention are jointly associated with autism. Autism Res*. 6(1): 11–22.
28. Russo N, Nicol T, Trommer B, Zecker S, Kraus N (2009): Brainstem transcription of speech is disrupted in children with autism spectrum disorders. *Dev Sci*. 12(4): 557–67.
29. Garreau B, Barthelemy C, Martineau J, Bruneau N, Lelord G (1985): [electrophysiologic aspects of infantile autism]. *Encephale*. 11(4): 145–55.
30. Sohmer H, Student M (1978): Auditory nerve and brain-stem evoked responses in normal, autistic, minimal brain dysfunction and psychomotor retarded children. *Electroencephalogr Clin Neurophysiol*. 44(3): 380–88.
31. Student M, Sohmer H (1978): Evidence from auditory nerve and brainstem evoked responses for an organic brain lesion in children with autistic traits. *J Autism Child Schizophr*. 8(1): 13–20.
32. Rosenhall U, Nordin V, Brantberg K, Gillberg C (2003): Autism and auditory brain stem responses. *Ear Hear*. 24(3): 206–14.
33. Tanguay PE, Edwards RM, Buchwald J, Schwafel J, Allen V (1982): Auditory brainstem evoked responses in autistic children. *Arch Gen Psychiatry*. 39(2): 174–80.
34. Courchesne E, Courchesne RY, Hicks G, Lincoln AJ (1985): Functioning of the brain-stem auditory pathway in non-retarded autistic individuals.

*Electroencephalogr Clin Neurophysiol*. 61(6): 491–501.
35. Screening Data |Early Hearing Detection and Intervention (EHDI) | NCBDDD | CDC. https://ehdidash.cdc.gov/IAS/dataviews/view?viewId=26
36. EHDI Programs | Hearing Loss | NCBDDD | CDC. http://www.cdc.gov/ncbddd/hearingloss/ehdi-programs.html
37. NCHAM: State Grants. http://www.infanthearing.org/stategrants/2011-archive.php
38. 2001 CDC State EHDI Grant: North Carolina Abstract & Narrative. http://www.infanthearing.org/stategrants/cdc2001/cdc2001_northcarolina.html
39. Harrison W, Goodman D (2015): Epidemiologic trends in neonatal intensive care, 2007-2012. *JAMA Pediatr*. 169(9): 855–62.
40. Dabbous AO (2012): Characteristics of auditory brainstem response latencies in children with autism spectrum disorders. *Audiol Med*. 10(3): 122–31.
41. Miron O, Ari-Even Roth D, Gabis LV, Henkin Y, Shefer S, Dinstein I, *et al. (2015): Prolonged auditory brainstem responses in infants with autism. Autism Res*
42. Wong V, Wong SN (1991): Brainstem auditory evoked potential study in children with autistic disorder. *J Autism Dev Disord*. 21(3): 329–40.
43. Sersen EA, Heaney G, Clausen J, Belser R, Rainbow S (1990): Brainstem auditory-evoked responses with and without sedation in autism and down's syndrome. *Biol Psychiatry*. 27(8): 834–40.
44. Ornitz EM, Mo A, Olson ST, Walter DO (1980): Influence of click sound pressure direction on brainstem responses in children. *Audiology*. 19(3): 245–54.
45. Gillberg C, Rosenhall U, Johansson E (1983): Auditory brainstem responses in childhood psychosis. *J Autism Dev Disord*. 13(2): 181–95.
46. Roth DA, Muchnik C, Shabtai E, Hildesheimer M, Henkin Y (2012): Evidence for atypical auditory brainstem responses in young children with suspected autism spectrum disorders. *Dev Med Child Neurol*. 54(1): 23–29.
47. Tas A, Yagiz R, Tas M, Esme M, Uzun C, Karasalihoglu AR (2007): Evaluation of hearing in children with autism by using teoae and abr. *Autism*. 11(1): 73–79.
48. Ververi A, Vargiami E, Papadopoulou V, Tryfonas D, Zafeiriou D (2015): Brainstem auditory evoked potentials in boys with autism: still searching for the hidden truth. *Iran J Child Neurol*. 9(2): 21–28.
49. Tanguay PE, Edwards RM (1982): Electrophysiological studies of autism: the whisper of the bang. *J Autism Dev Disord*. 12(2): 177–84.
50. Tharpe AM, Bess FH, Sladen DP, Schissel H, Couch S, Schery T (2006): Auditory characteristics of children with autism. *Ear Hear*. 27(4): 430–41.
51. Russo NM, Skoe E, Trommer B, Nicol T, Zecker S, Bradlow A, *et al. (2008): Deficient brainstem encoding of pitch in children with autism spectrum disorders. Clin Neurophysiol*. 119(8): 1720–31.
52. Rosenblum SM, Arick JR, Krug DA, Stubbs EG, Young NB, Pelson RO (1980): Auditory brainstem evoked responses in autistic children. *J Autism Dev Disord*. 10(2): 215–25.
53. Fujikawa-Brooks S, Isenberg AL, Osann K, Spence MA, Gage NM (2010): The effect of rate stress on the auditory brainstem response in autism: a preliminary report. *Int J Audiol*. 49(2): 129–40.
54. Magliaro FC, Scheuer CI, Assumpção Júnior FB, Matas CG (2010): Study of auditory evoked potentials in autism. *Pro Fono*. 22(1): 31–36.
55. Rumsey JM, Grimes AM, Pikus AM, Duara R, Ismond DR (1984): Auditory brainstem responses in pervasive developmental disorders. *Biol Psychiatry*. 19(10): 1403–18.
56. Grillon C, Courchesne E, Akshoomoff N (1989): Brainstem and middle latency auditory evoked potentials in autism and developmental language disorder. *J Autism Dev Disord*. 19(2): 255–69.